\documentclass[12pt]{iopart}
\usepackage[dvips]{graphicx} 
\begin{document}

\title[New Insight in Physics]
{\begin{center}
\bfseries\Large UNIVERSALITY
\end{center}}
\vskip 1.0cm
\author{Emil Marinchev}

\address{ Technical University of Sofia

Physics Department

8, Kliment Ohridski St.

Sofia-1000, BG

e-mail: emar@tu-sofia.acad.bg}
\vskip 2.0 cm
\begin{abstract}
This article is an attempt for a new vision of the basics of Physics,
and of Relativity, in particular. A new generalized principle of inertia is
proposed, as an universal principle, based on universality of the conservation laws, 
not depending on the metric geometry used. The second and the third principles of 
Newton's mechanics are interpreted as logical consequences. The generalization of the
classical principle of relativity  made by Einstein
as the most basic postulate in the Relativity is criticized as logically not
well-founded. A new theoretical scheme is proposed based on two basic
principles:

\;1.The principle of universality of the conservation laws, and

\;2.The principle of the universal velocity.  

It is well- founded with examples of different fields of physics.

\end{abstract}

\skip 1.0 cm
Comments: 5 pages, 1 figure

Subj-class: General Physics


\maketitle

\section{Introduction}
Theoretical physics describes geometrically the motion of matter,
substituting real physical  objects by geometrical objects (models). The
geometry that is usually used depends on the distribution of matter and its
motion. In small velocities Euclidian geometry is used (Newtonian mechanics).
In velocities comparable to the speed of light pseudo-Euclidian geometry is
used.  Pseudo-Riemannian geometry is used in General Relativity when material
structures with high energy-momentum density are considered.  The following
geometrical objects are used for the quantitative description of physical
phenomena:  scalars, vectors, poly-vectors, differential forms, tensors etc. Since the essential
nature of the physical phenomena do not depend on the reference frame,
the corresponding geometrical objects should be absolute in the sense, that
they do not depend on the choice of the coordinate system and frame used, although 
that they may have different presentation in the various ones.  By their very
nature the physical laws are universal, so their geometrical equivalents
should be absolute too.

A full description of any physical phenomenon includes kinematical and
dynamical (in particular, statics) parts.  There is relation between these
two parts because they deal with the same physical phenomenon.  The
kinematics describes the motion of the system considered making use of
concepts like trajectory, velocity, acceleration, or with their graphical
presentation. The main feature of the dynamical description is making use of
conservative quantities like momentum, angular momentum, energy, etc.  In
presence of interaction the universal conserved quantities momentum, angular
momentum and energy are exchanged with or without exchange of mass
(particles) and other physical quantities like electrical charge and/or other
charges.

If a physical system does not interact with other (external) objects it is a
closed system and it is natural to describe its motion with conservative
quantities. The idea of a closed (isolated) system is a very useful one
because the dynamical part of the description is very simple.  As
we know, Newton built his mechanics on three principles. But aren't they more
than necessary? The most significant principle, in my view, is the first one,
the principle of inertia.  The idea of this paper is to present an appropriate 
generalization of this principle to an universal one, i.e. not depending on the 
metric geometry used,  and   the other two principles of Newton  to follow  as 
logical  consequences of it.

\section{Universality - Generalized Principle of Inertia}
Our generalized principle of inertia may be formulated as follows:

{\bf Reference frames in which the physical systems conserve their state of
motion, if they do not interact with other objects, are universal}.

Let's give some clarifying comments. The quality of "conservation of state of
motion" we call inertia, and we characterize it by a set of conservative
quantities - momentum ${\mathbf p}$, angular momentum ${\mathbf L}$, energy $E$ and mass $m$. 
The generalized principle of inertia automatically includes the inertial rotation.
Traditionally, in physics, it is accepted that inertia is determined only by
the scalar quantity mass. The first principle of Newton says that the state
of uniform motion along a straight line is conserved.  Obviously the
direction of motion is conserved and it could not be described by a scalar
quantity.  Photons, although they have no mass, they have momentum and move by inertia 
if are not subject to interference. The inertial motion is with constant values of the
 conservation quantities. Clearly, the universal reference frames are a generalization of 
the usual inertial frames. 

\begin{equation}
\eqalign{
\mathbf {p=const}\\
\mathbf {L=const}\\
E=const\\
m=const}
\end{equation}
If there is interaction the state of motion is changed and the conservation
quantities are changed too. The rate of change of the conservation quantities
defines quantitatively the interaction with the surrounding objects, i.e.
physical quantities: force ${\mathbf F}$, torque ${\mathbf M}$, work $A$ 
and reactive force ${\mathbf F}_{\rm R}$.
\begin{equation}
\eqalign{
\mathbf {\dot p=F}\\
\mathbf {\dot L=M=r\times F}\\
\Delta E=A\\
\dot m\mathbf {u=F} \rm _R}
\end{equation}
In exchange of momentum between two bodies one of them recoils and the
other one  accepts momentum, and as a result action and reaction are
equal in magnitude and opposite in direction. According to the Newton's third
principle the horse and the cart are pulling each other, but the leading part
of the horse is obvious. In our approach to dynamics the laws of conservation
are the leading ones, not the forces of interaction.

\section{Examples}
1. It is not by chance that the first principle of thermodynamics is
an application of the law of energy conservation in thermodynamic systems and
the transport phenomena (the viscosity, thermal conductivity and diffusion)
are associated with exchange of momentum, energy and particles in micro
level. The main equation of the molecular kinetic theory and the consequences
of it are easily defined if we look at the pressure as energy density
 and in the same time as  bidirectional exchange of momentum
in arbitrary direction in the frame of the degrees of freedom allowed:  
\begin{equation*} p=2n{\bar
\varepsilon}/i=2n{\bar \varepsilon_1}=nkT\;\; \Rightarrow \; \;  {\bar
\varepsilon_1}=\frac12kT, 
\end{equation*}
$n$ is the number of molecules in unit volume,     
$\bar \varepsilon$ is the mean energy of one molecule, $i$ is the number of degrees of freedom, 
$T$ is the temperature in K, $k$ is the Boltzmann constant.

2. A typical reactive force is the lift force in aerodynamics
${\mathbf F}_{\rm R}$, Figure 1.

\begin{figure}[h]
\begin{center}
\includegraphics[height=1in,width=3in]{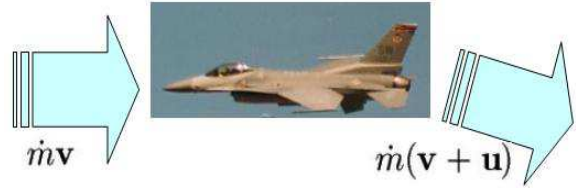}
\end{center}
\caption{Lift \; \; \; \;\; ${\mathbf {F} _{\rm R}={\dot m} \mathbf {u}={\dot m} (\mathbf {v+u})-{\dot m} \mathbf {v=-F}}$}
\end{figure}

3. In the theory of relativity the laws of conservation are united in a
general law of conservation by the four-vector of the momentum $\bf p$. In
the general theory of relativity the curvature of time-space is used instead of
gravitational interaction. The gravitational interaction in the Newtonian
sense is missing. The motion of cosmic objects is in harmony with the
conservation laws and is along the extreme line:

\begin{equation}
\eqalign{
\mathbf {p}=(E,\vec p)=m\mathbf {u}, \;
\; m\equiv|\mathbf {p}|=\rm E_0, \; \; |\mathbf {u}|=c=1\\
\mathbf {\dot p=F}, \; \; \mathbf {\dot p=(\nabla p).u=\nabla_u p=\nabla.(pu)=\nabla.T}\\
\mathbf {\dot p=(\nabla p).u=\nabla_u p=\nabla.(pu)=0},\;
\;  if \; \mathbf {F=0,}}
\end{equation}

$\mathbf{T}$ is the tensor of energy-momentum, $\mathbf{pu \equiv p \otimes u},
\;\;\mathbf{\nabla u \equiv \nabla \otimes u}$.

The local motion is in a straight line and the time-space is flat.

The principle of relativity is the first postulate in the special and in the
general theory of relativity which probably has given their names.  From our
point of view  the development of these theories could be done without the
principle of relativity, as it is in Newtonian mechanics.  The second
postulate of Einstein can be generalized into the following universal
principle:
 
{\bf There exists maximal possible universal velocity common for all
reference frames} $c=3.10^8$ $m/s$ ($c=1$). {\bf The velocity of light in vacuum is
equivalent to it.}
 
As for the usage of the concept of relativity, in our
opinion the concept of UNIVERSALITY is much more adequate because of the
universality of the conservation laws, of the principle of the universal
velocity, and the universality of gravitation.

4. In electromagnetism, the rate of change of the momentum of the $q$-charged
particle defines quantitatively the interaction with the
external electromagnetic field, i.e. the Lorentz force ${\bf F}_{\rm L}$.
\begin{equation}
\eqalign{
\dot {\vec p}=\mathbf {F} _{\rm L} = q (\mathbf {E+v\times B})\\
\mathbf {\dot p=(\nabla p).u}=q\mathbf {F.u},}
\end{equation}
here $\bf F$ is the tensor of the electromagnetic field.

5. In the microworld, the universality of conservation laws and the principle of
universal velocity demonstrate their universality even more substantially,
and most frequently they give the only possibility to explain these
microphenomena.

We would like to note that this approach of using conservation laws as
dynamics generating rules has been used constructively in [1,2] where
a nonlinear generalization of Maxwell equations giving more realistic
description of the electromagnetic phenomena is achieved.

\section{Universality applied to Gravitation}
The time-space tells the matter how to move by inertia according to the laws of
 conservation.

\begin{equation}
\eqalign{
\mathbf {\dot p=(\nabla p).u=\nabla_u p=\nabla.(pu)=0},\;
\;  if \; \mathbf {F=0}}.
\end{equation}
Obviously between time-space and matter there is no exchange of
energy-momentum.  The changes of time-space should be characterized in a
similar way to the motion of matter by a symmetric tensor of second rank
with covariant divergence equal to zero, e.g. the tensor $\mathbf G$ of Hilbert-Einstein:

\begin{equation}
\eqalign{
\mathbf {\nabla.G=\nabla.(R-\mathit {\frac12 R} g)=0=\nabla.T}\;\;\Rightarrow \\
\mathbf{G=R}-
\mathit {\frac12 R} \mathbf {g}=\kappa \mathbf{T},
\;\;\; \kappa=8\pi \gamma,}
\end{equation}
where $\mathbf T$ is the tensor of the density and the flux of energy-momentum.

In the theoretical scheme considered here the only concept of mass was used: $m = |\mathbf p|$, 
and the principle of equivalence automatically follows. The time-space curvature, 
i.e. the gravitation, is determined by the matter distribution and its motion $\mathbf T$, 
and only in the static case without other fields  by the mass distribution $\mathbf p = (m,\mathbf 0)$. 

Finally, I'd like to note that this vision on physics is inspired by
discussions with S.Donev. I kindly acknowledge his critical views during the
discussions, which contributed greatly to clarify some difficult issues in
the subject.

\vskip 2.5cm
{\bf References}
\vskip 0.5cm
[1].  {\bf Donev, S., Tashkova, M}., {\it Energy-momentum Directed
Nonlinearization of Maxwell's Equations in the Case of Continuous Media},
Proc. R. Soc.  Lond.  A, 443 (1995), 281-291

[2]. {\bf Donev, S.}, {\it Parallel Objects and Field Equations}, LANL
e-print: math-ph/0205046

\end{document}